\documentclass[12pt]{article}

\usepackage{graphicx}

\usepackage{bbm}

\newcommand{\be}{\begin{equation}}
\newcommand{\ee}{\end{equation}}
\newcommand{\ba}{\begin{eqnarray}}
\newcommand{\ea}{\end{eqnarray}}

\newcommand{\mnu}{\mathcal{M}_\nu}

\newcommand{\lesssim}{\:\mbox{\raisebox{-3pt}{$\stackrel%
{\displaystyle <}{\sim}$}}\:}
\newcommand{\gtrsim}{\:\mbox{\raisebox{-3pt}{$\stackrel%
{\displaystyle >}{\sim}$}}\:}

\begin{document}

\title{\normalsize \hfill UWThPh-2006-22 \\[8mm]
\LARGE Predictions of an $A_4$ model \\
with a five-parameter neutrino mass matrix}

\author{
\setcounter{footnote}{2}
L.\ Lavoura$^{(1)}$\thanks{E-mail: balio@cftp.ist.utl.pt}
\ and H.\ K\"uhb\"ock$^{(2)}$\thanks{E-mail: helmut.kuehboeck@gmx.at}
\\*[4mm]
$^{(1)}$ \small Universidade T\'ecnica de Lisboa
and Centro de F\'\i sica Te\'orica de Part\'\i culas \\
\small Instituto Superior T\'ecnico, 1049-001 Lisboa, Portugal
\\*[3mm]
$^{(2)}$ \small Institut f\"ur Theoretische Physik, Universit\"at Wien \\
\small Boltzmanngasse 5, A--1090 Wien, Austria \\*[3.6mm]
}

\date{4 October 2006}

\maketitle

\begin{abstract}
We consider a model with an $A_4$ flavour symmetry,
recently proposed by E.\ Ma~\cite{ma}, and
make a numerical study,
through scatter plots,
of its neutrino mass matrix.
\end{abstract}

\vspace*{10mm}

The neutrino mass matrix $\mnu$
is defined by the neutrino mass term
\be
\mathcal{L}_{\nu\, \rm{mass}} =
\frac{1}{2}\, \nu_L^T C^{-1} \mnu\, \nu_L
+ \rm{H.c.},
\ee
where $C$ is the Dirac--Pauli charge-conjugation matrix
and $\nu_L = \left( \nu_{eL},\, \nu_{\mu L},\, \nu_{\tau L} \right)^T$
is the column vector
of the three weak-eigenstate left-handed neutrino fields.
The matrix $\mnu$ is symmetric and is diagonalized as
\be
U^T \mnu\, U = \mbox{diag} \left( m_1,\, m_2,\, m_3 \right),
\ee
where the lepton mixing matrix $U$
is unitary and the neutrino masses $m_{1,2,3}$
are real and non-negative.
The matrix $U$ may be parametrized as
\be
U =
P_F \left( \begin{array}{ccc}
c_{13} c_{12} &
c_{13} s_{12} &
s_{13} e^{- i \delta} \\
- c_{23} s_{12} - s_{23} s_{13} c_{12} e^{i \delta} &
c_{23} c_{12} - s_{23} s_{13} s_{12} e^{i \delta} &
s_{23} c_{13} \\
- s_{23} s_{12} + c_{23} s_{13} c_{12} e^{i \delta} &
s_{23} c_{12} + c_{23} s_{13} s_{12} e^{i \delta} &
- c_{23} c_{13}
\end{array} \right) P_M,
\label{U}
\ee
where
\ba
P_F &=& \mbox{diag} \left(
e^{i \vartheta_e},\, e^{i \vartheta_\mu},\, e^{i \vartheta_\tau}
\right),
\\
P_M &=& \mbox{diag} \left(
e^{i \Theta / 2},\, 1,\, e^{i \Omega / 2}
\right)
\ea
are diagonal unitary matrices.
In equation~(\ref{U}),
$c_{ij} = \cos{\theta_{ij}}$
and $s_{ij} = \sin{\theta_{ij}}$
for $ij = 23, 13, 12$.
There are nine observables:
the three masses $m_{1,2,3}$,
the three mixing angles $\theta_{23,13,12}$,
the Dirac phase $\delta$,
and the Majorana phases $\Theta$ and $\Omega$.\footnote{The
phases $\vartheta_\alpha$ ($\alpha = e, \mu, \tau)$ are not observable
since they may be eliminated
through a rephasing of the charged-lepton fields.}
Correspondingly,
the matrix $\mnu$ has in general nine parameters:
the moduli of its six independent matrix elements,
and three phases invariant under $\mnu \to X \mnu\, X$,
where $X$ is an arbitrary diagonal unitary matrix.

Recently~\cite{ma},
E.\ Ma proposed a model,
based on the seesaw mechanism and on a $A_4$ symmetry group,
wherein $\mnu$ may be parametrized as
\be
\mnu = \left( \begin{array}{ccc}
f a^2 & a b & a c \\
a b & f b^2 & b c \\
a c & b c & f c^2
\end{array} \right),
\label{mnu}
\ee
where $f$ is dimensionless
while $a$,
$b$,
and $c$ have dimension $M^{1/2}$.
We may,
without loss of generality,
assume $a$,
$b$,
and $c$ to be real and non-negative:
their phases may be absorbed
through appropriate rephasings of the $\nu_{\alpha L}$.
Ma's model is interesting
since it has the potential to be quite predictive:
it has only five parameters---$a$,
$b$,
$c$,
and the modulus and phase of $f$.
This model has as many parameters as the two-texture-zero models
of Frampton, Glashow, and Marfatia~\cite{FGM}
and of Lavoura~\cite{Lavoura}.
It is also remarkable that the form of the mass matrix~(\ref{mnu})
is invariant under the renormalization-group evolution
\be
\mnu \left( \mu \right) =
I \left( \mu, \mu_0 \right)
\mnu \left( \mu_0 \right)
I \left( \mu, \mu_0 \right),
\ee
where $I \left( \mu, \mu_0 \right)$ is a diagonal matrix;
indeed,
the diagonal matrix elements of $I \left( \mu, \mu_0 \right)$
may be absorbed through redefinitions of $a$,
$b$,
and $c$
while keeping the form of $\mnu$ in~(\ref{mnu}) unchanged.

When $b=c$ the matrix~(\ref{mnu}) becomes $\mu$--$\tau$ symmetric,
leading therefore to $\theta_{23} = \pi/4$ and $\theta_{13} = 0$.
This is allowed by the experimental data---see below.
But,
it is important to study in detail the model~(\ref{mnu}) for $\mnu$
in order to see how far away from that limiting case
$\theta_{23} = \pi/4$, $\theta_{13} = 0$
that model allows one to go.
That's the study that we undertake in this letter.
Since it seems difficult to proceed analytically,
we have done our study in a purely numerical fashion,
by constructing appropriate scatter plots.

The mass matrix~(\ref{mnu}) is characterized by the relations
\be
\begin{array}{rcl}
{\displaystyle
\frac{\left( \mnu \right)_{ee}}{\left( \mnu \right)_{\mu\mu}}
}
&=&
{\displaystyle
\left[
\frac{\left( \mnu \right)_{e\tau}}
{\left( \mnu \right)_{\mu\tau}}
\right]^2,
}
\\*[3mm]
{\displaystyle
\frac{\left( \mnu \right)_{ee}}{\left( \mnu \right)_{\tau\tau}}
}
&=&
{\displaystyle
\left[
\frac{\left( \mnu \right)_{e\mu}}
{\left( \mnu \right)_{\mu\tau}}
\right]^2,
}
\end{array}
\label{scaling1}
\ee
which should be compared to the analogous ``scaling'' relations,
recently suggested by Mohapatra and Rodejohann (MR),
\be
\frac
{\left( \mnu \right)_{e\mu}}
{\left( \mnu \right)_{e\tau}}
=
\frac
{\left( \mnu \right)_{\mu\mu}}
{\left( \mnu \right)_{\mu\tau}}
=
\frac
{\left( \mnu \right)_{\mu\tau}}
{\left( \mnu \right)_{\tau\tau}}\, .
\label{scaling2}
\ee
Both the scaling relations~(\ref{scaling1})
and~(\ref{scaling2}) are invariant
under the renormalization-group evolution,
and both lead to mass matrices parametrized by four moduli and one phase.
However,
while the MR scaling relations\footnote{Models
leading to the MR scaling relations had already been proposed
before the MR paper~\cite{lavoura,grimus}.}
lead to the clear-cut predictions
$m_3 = 0$ and $\theta_{13} = 0$,\footnote{This is equivalent
to four predictions,
since $m_3 = 0$ implies that the Majorana phase $\Omega$ is meaningless,
and $\theta_{13} = 0$ implies that the Dirac phase $\delta$ is meaningless.}
we shall find in the present letter
that the scaling relations~(\ref{scaling1})
lead to much less well-defined predictions,
largely because of the experimental indeterminacy
of the Majorana phases $\Theta$ and $\Omega$.

We have used in our numerical study
the following data on the mixing angles:
\be
0.30 < \tan^2{\theta_{12}} < 0.61,
\label{solar}
\ee
and
\ba
\cos^2{2 \theta_{23}} &<& 0.10
\label{atm} \\
\sin^2{\theta_{13}} &<& 0.047.
\label{Ue3}
\ea
These are the $3 \sigma$ bounds
derived in~\cite{maltoni} from the data.
Notice that,
experimentally,
there is no preferred value for $\sin^2{\theta_{13}}$:
there is only an upper bound on it.
As for $\theta_{23}$,
its experimentally preferred value is $\pi / 4$
and it is equally probable that it is higher or lower than $\pi / 4$;
for this reason,
$\cos^2{2 \theta_{23}}$ seems to be the appropriate observable.
For the neutrino masses we have taken---see~\cite{maltoni}---
\be
0.022 <
\frac{m_2^2- m_1^2}{\left| m_3^2- m_1^2 \right|}
< 0.065.
\label{ratio}
\ee
The sign of $m_3 - m_1$ is unknown;
mass spectra with $m_3 > m_1$ are called ``normal'',
those with $m_3 < m_1$ are dubbed ``inverted''.

We have made extensive scans of the space parametrized by $b/a$,
$c/b$,
$|f|$,
and $\arg{f}$.
Whenever the resulting observables
agreed with the data~(\ref{solar})--(\ref{ratio}),
we would fix the overall scale of the neutrino masses by using
\be
m_2^2- m_1^2 =
8.1 \times 10^{-5}\, \mathrm{eV}^2.
\ee
Our results are given by scatter plots 1--8.
We have found,
numerically,
that a normal mass spectrum is obtained when $c \ge b > a$,
and an inverted one when $c \le b < a$.
In both cases,
good fits of the observables
are only obtained when $\arg{f}$ is very close or equal to $\pi$.

Our main plots are~1 and~2,
in which one sees $\sin^2{\theta_{13}}$
plotted against $\cos^2{ 2 \theta_{23}}$.
The situation $\sin^2{\theta_{13}} = \cos^2{ 2 \theta_{23}} = 0$
is always allowed,
since it corresponds to $b=c$.
In general,
any value of $\cos^2{ 2 \theta_{23}}$ obeying the bound~(\ref{atm}),
and of $\sin^2{\theta_{13}}$ obeying the bound~(\ref{Ue3}),
is allowed by Ma's model;
none of these observables may be zero unless the other one vanishes too.
One also sees in figures~1 and~2 that the model generally predicts
\be
\sin^2{\theta_{13}} \lesssim \frac{\cos^2{2 \theta_{23}}}{2}\, .
\label{bound}
\ee
When the neutrino mass spectrum is inverted,
there is also a distinctive bound
\be
\sin^2{\theta_{13}} \gtrsim \frac{15 \cos^2{2 \theta_{23}}}{100}
\quad \mbox{(inverted spectrum)}.
\label{bound2}
\ee

We gather from plots~1 and~2 that,
in the case of a normal spectrum,
a larger area of the $\cos^2{ 2 \theta_{23}}$--$\sin^2{\theta_{13}}$ plane
is allowed than in the case of an inverted spectrum.
Ma's model seems to prefer in general
a normal neutrino mass spectrum.
Indeed,
while in that case any value for $\tan^2{\theta_{12}}$
in the range~(\ref{solar}) is allowed,
when the neutrino mass spectrum is inverted
$\tan^2{\theta_{12}}$ cannot be smaller than $0.5$.
This is depicted in plots~3 and~4,
respectively.

Almost-degenerate neutrinos
(the situation where $m_1^2 \gg \left| m_3^2 - m_1^2 \right|$)
are allowed both with normal and inverted mass spectra.
Consulting the plots~5 and~6,
we see that, in the case of a normal neutrino mass spectrum,
the lowest-mass neutrino may have an almost vanishing mass,
while,
when the mass spectrum is inverted,
the lowest neutrino mass must always be larger than $10^{-2}\, \mathrm{eV}$.

Experiments on neutrinoless double-beta decay
indirectly probe the modulus of the $ee$ matrix element of $\mnu$,
which we denote $\langle m \rangle_{ee}$.
This quantity is in Ma's model distinctly correlated
with the neutrino masses;
we see in plots~7 and~8 that $\langle m \rangle_{ee}$
is always extremely close to one third the average mass of the neutrinos.
This is due to the fact that good fits of the experimental observables
are in general only obtained when the ratios $b/a$ and $c/b$
are both very close to 1.

Summarizing,
we have analyzed numerically a neutrino mass matrix
which may be enforced through a renormalizable model,
has few parameters,
and is invariant under the renormalization-group evolution.
We have found that that neutrino mass matrix
has less predictive power than initially guessed.
It is able to fit both normal and inverted neutrino mass spectra,
though in the latter case only if $\tan^2{\theta_{12}} \ge 0.5$.
The mass responsible for neutrinoless double-beta decay
is always very close to one-third the average neutrino mass.
The small parameters $\sin^2{\theta_{13}}$
and $\cos^2{2 \theta_{23}}$
may take any value in their experimentally allowed ranges,
but they always obey the bound~(\ref{bound}),
ans also the bound~(\ref{bound2})
if the neutrino mass spectrum is inverted.

\paragraph{Acknowledgements:}
We thank Walter Grimus for reading the manuscript
and making many useful suggestions.
The work of L.L.\ was supported by the Portuguese
\textit{Funda\c c\~ao para a Ci\^encia e a Tecnologia}
through the projects POCTI/FNU/37449/2001
and U777--Plurianual.

\newpage

\begin{figure}[ht]
\centering
\includegraphics[clip,height=100mm]{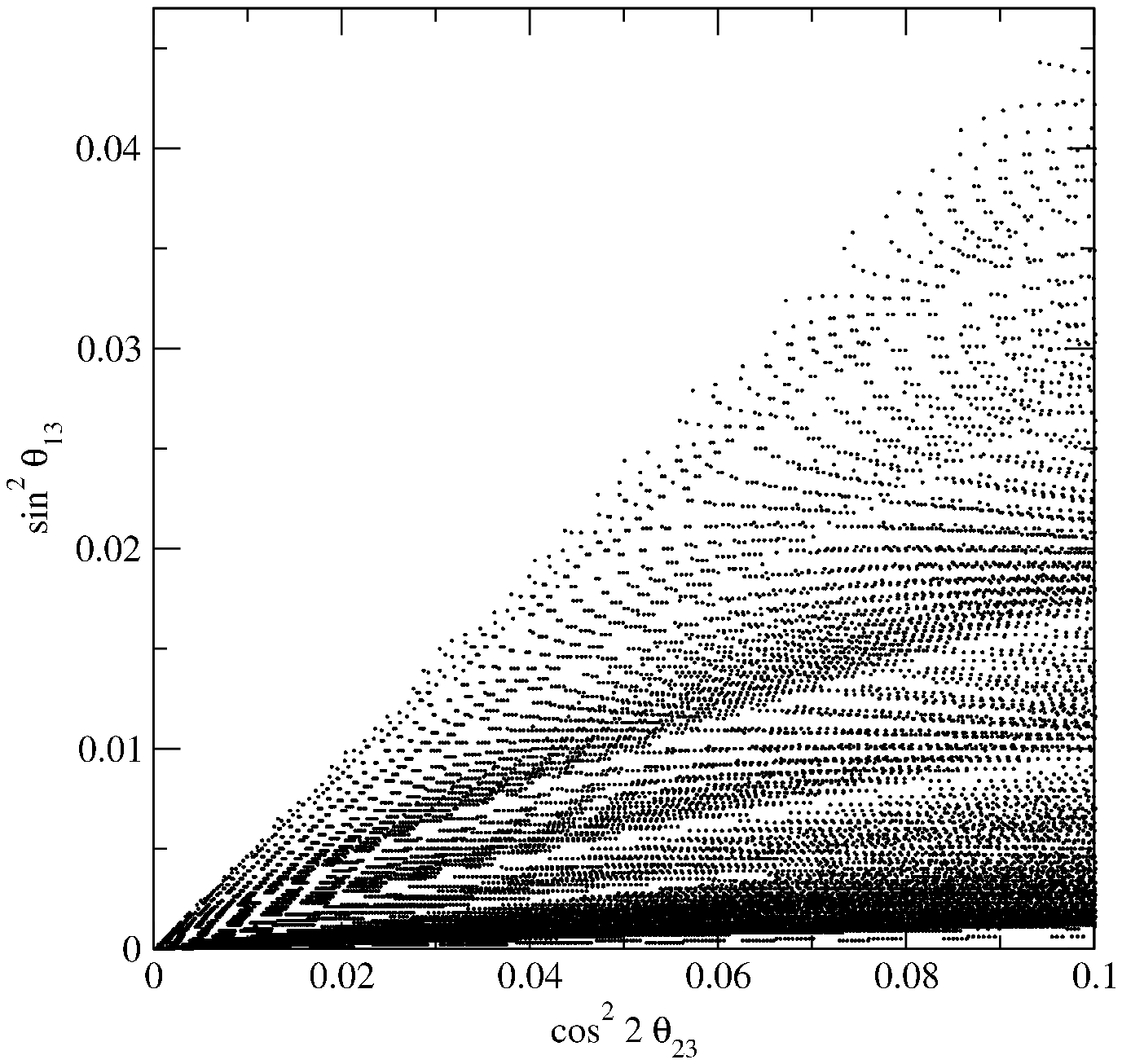}
\caption{Scatter plot of $\sin^2{\theta_{13}}$
as a function of $\cos^2{2 \theta_{23}}$
in the case of a normal neutrino mass spectrum.}
\label{fig1}
\end{figure}
\begin{figure}[hb]
\centering
\includegraphics[clip,height=100mm]{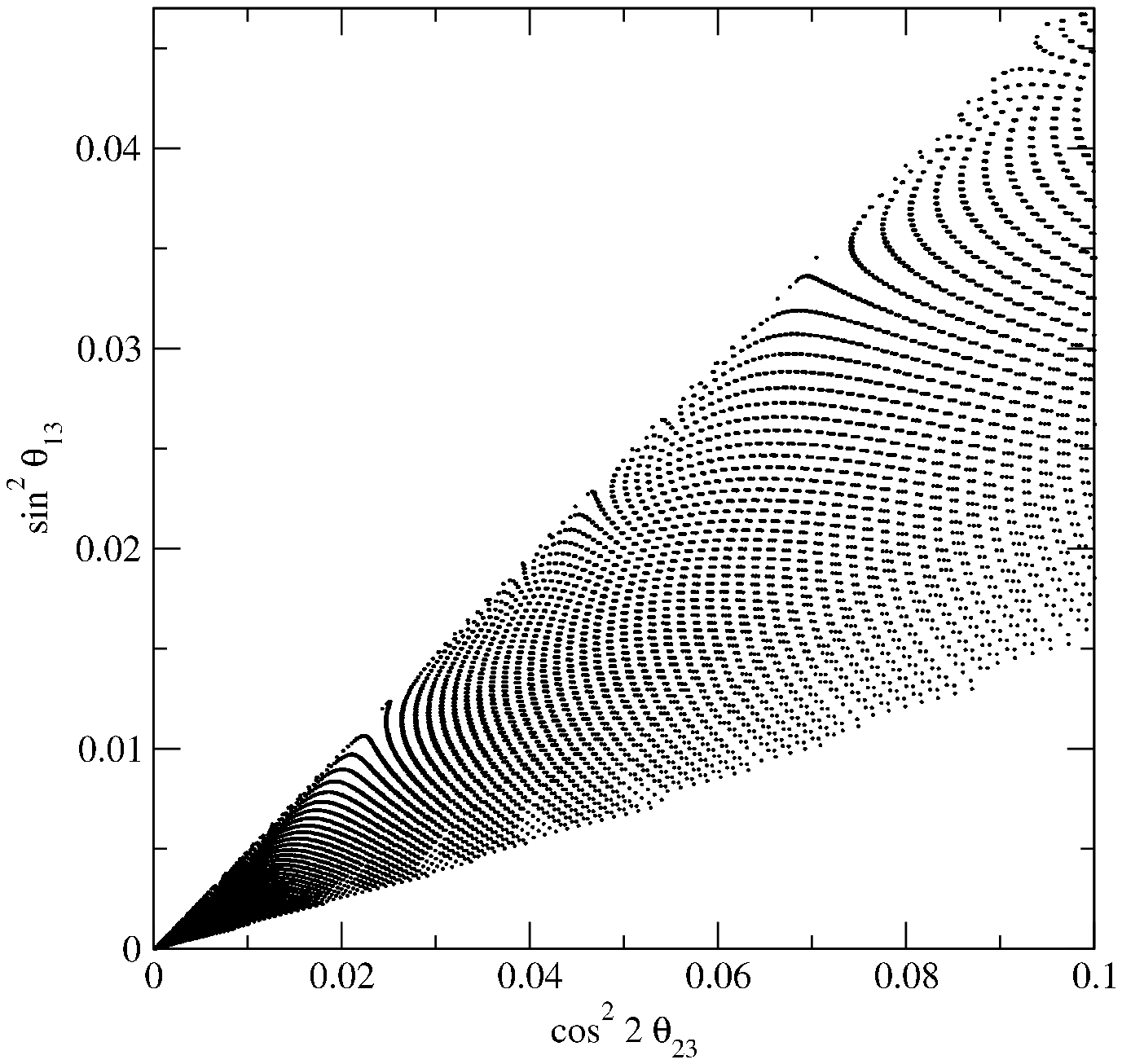}
\caption{The same as figure~1,
but in the case of an inverted neutrino mass spectrum.}
\label{fig2}
\end{figure}

\newpage

\begin{figure}[ht]
\centering
\includegraphics[clip,height=100mm]{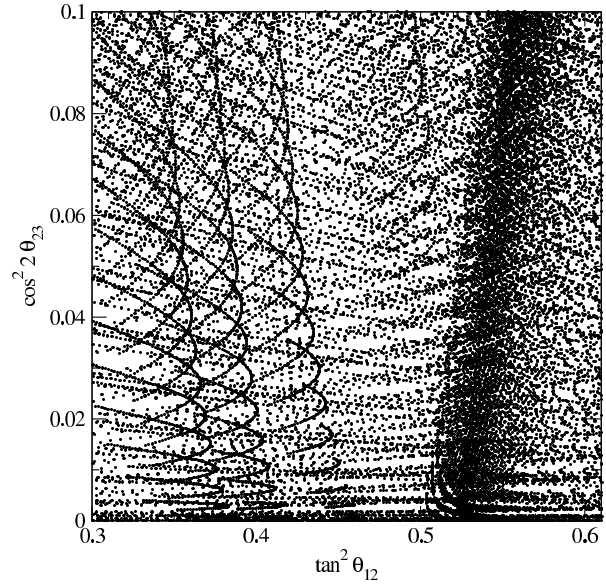}
\caption{Scatter plot of $\cos^2{2 \theta_{23}}$
as a function of $\tan^2{\theta_{12}}$
in the case of a normal neutrino mass spectrum.}
\label{fig3}
\end{figure}
\begin{figure}[hb]
\centering
\includegraphics[clip,height=100mm]{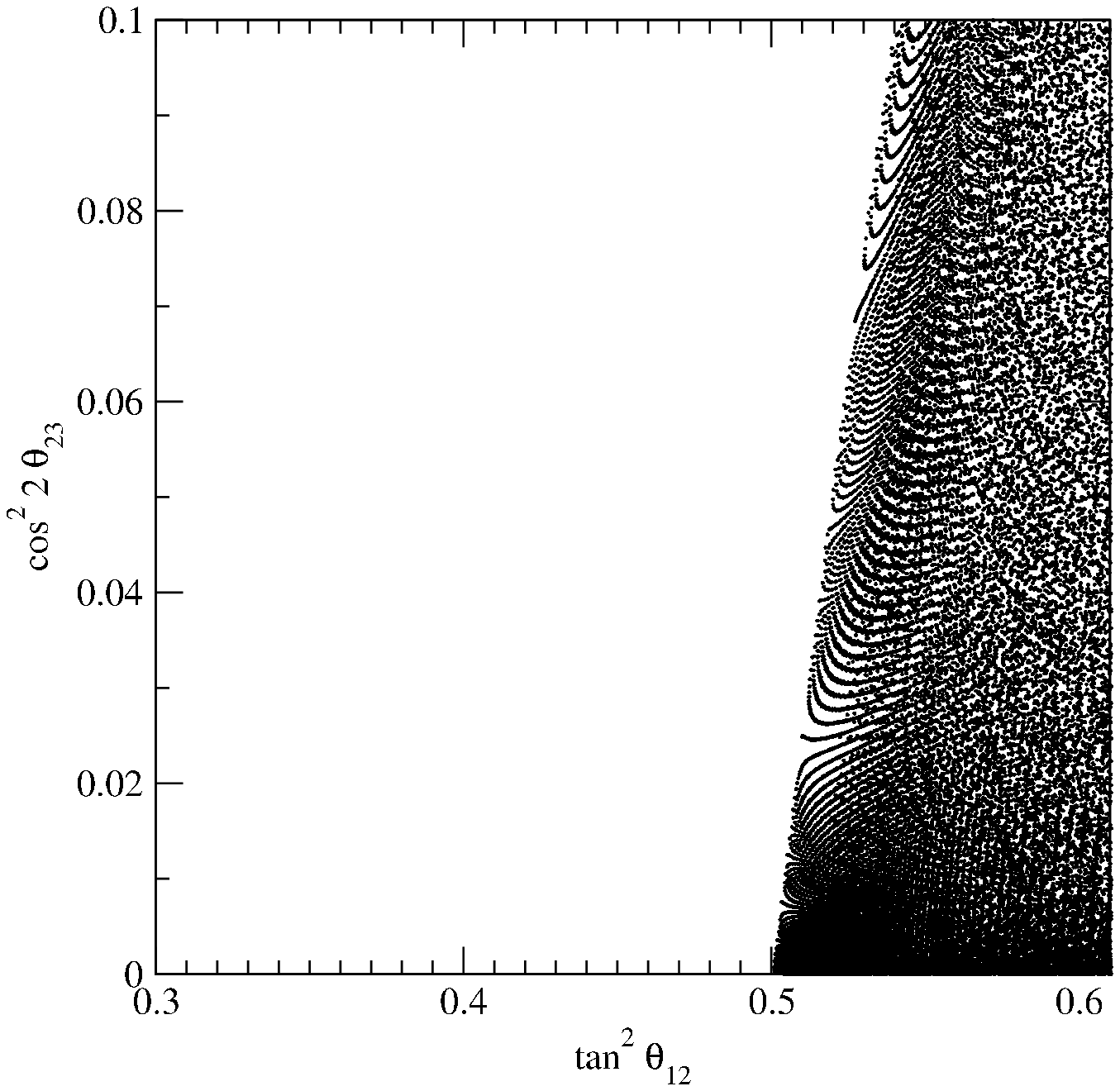}
\caption{The same as figure~3,
but in the case of an inverted neutrino mass spectrum.}
\label{fig4}
\end{figure}

\newpage

\begin{figure}[ht]
\centering
\includegraphics[clip,height=100mm]{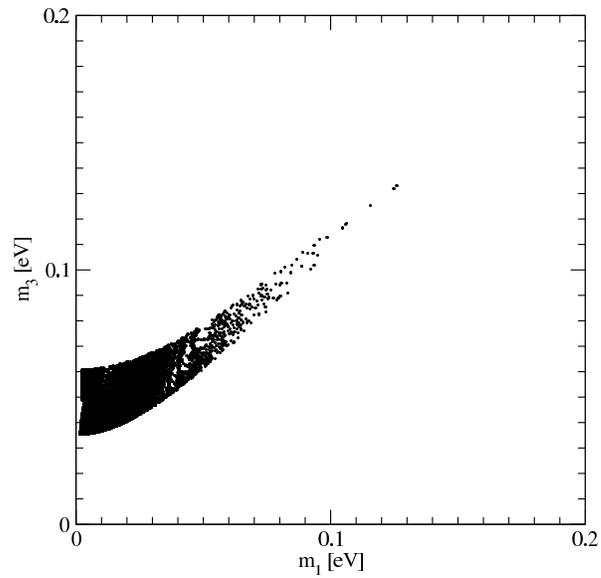}
\caption{Scatter plot of $m_3$ against $m_1$
in the case of a normal neutrino mass spectrum.}
\label{fig5}
\end{figure}
\begin{figure}[hb]
\centering
\includegraphics[clip,height=100mm]{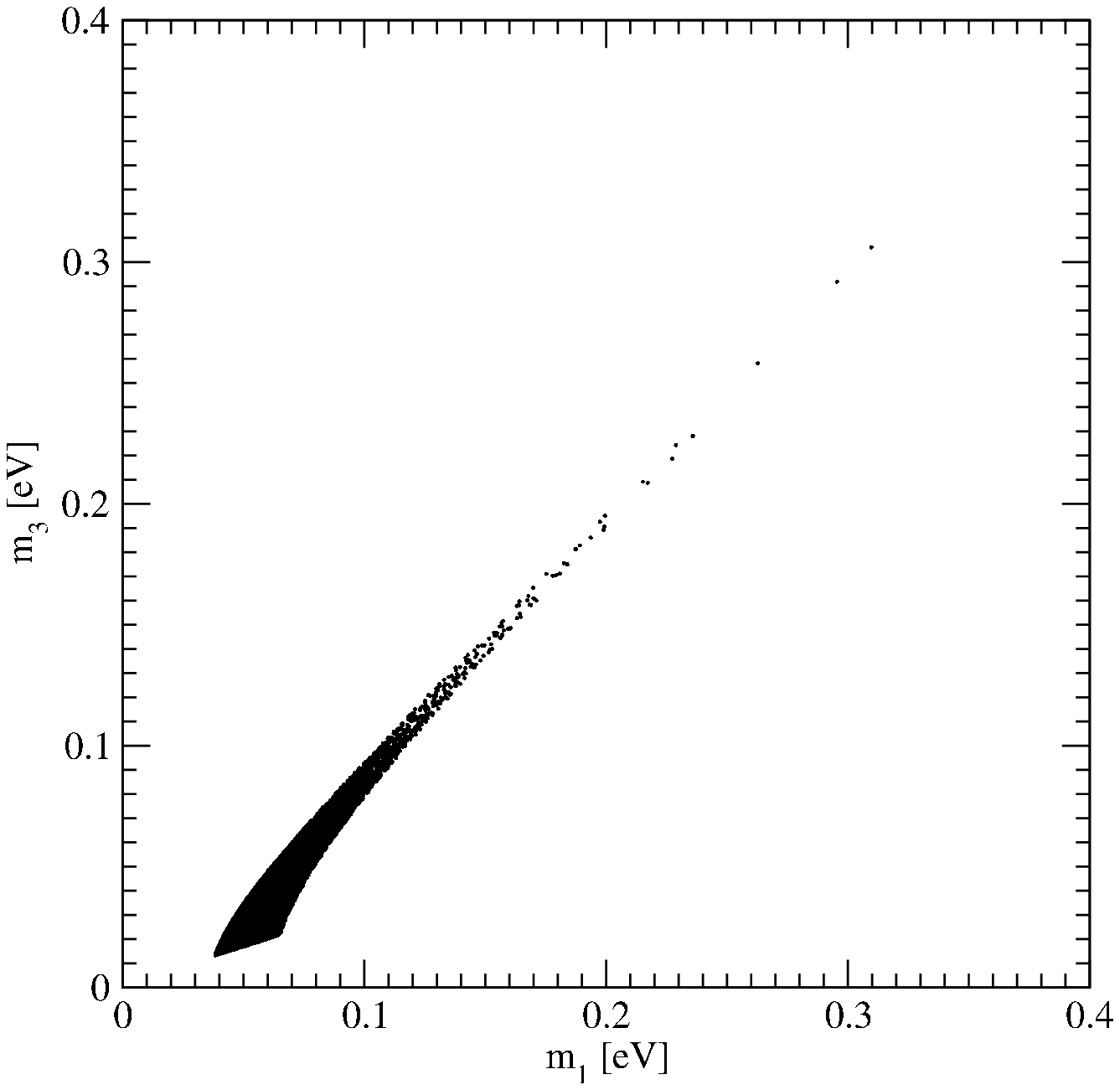}
\caption{The same as figure~5,
but in the case of an inverted neutrino mass spectrum.}
\label{fig6}
\end{figure}

\newpage

\begin{figure}[ht]
\centering
\includegraphics[clip,height=100mm]{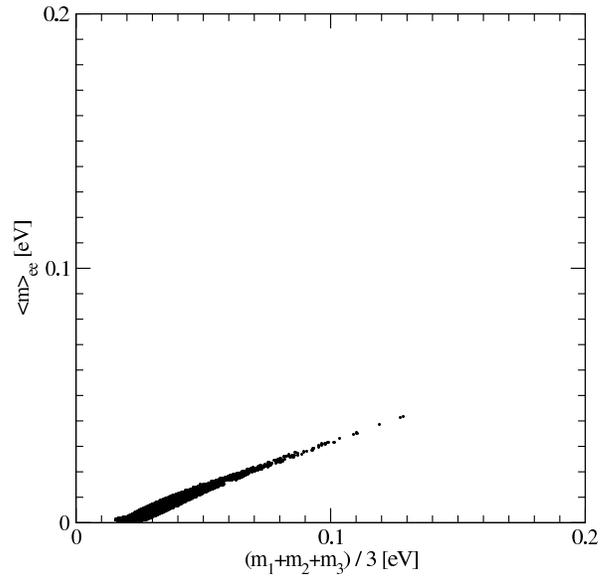}
\caption{The mass relevant for neutrinoless double-beta decay
as a function of the average mass of the neutrinos
in the case of a normal neutrino mass spectrum.}
\label{fig7}
\end{figure}
\begin{figure}[hb]
\centering
\includegraphics[clip,height=100mm]{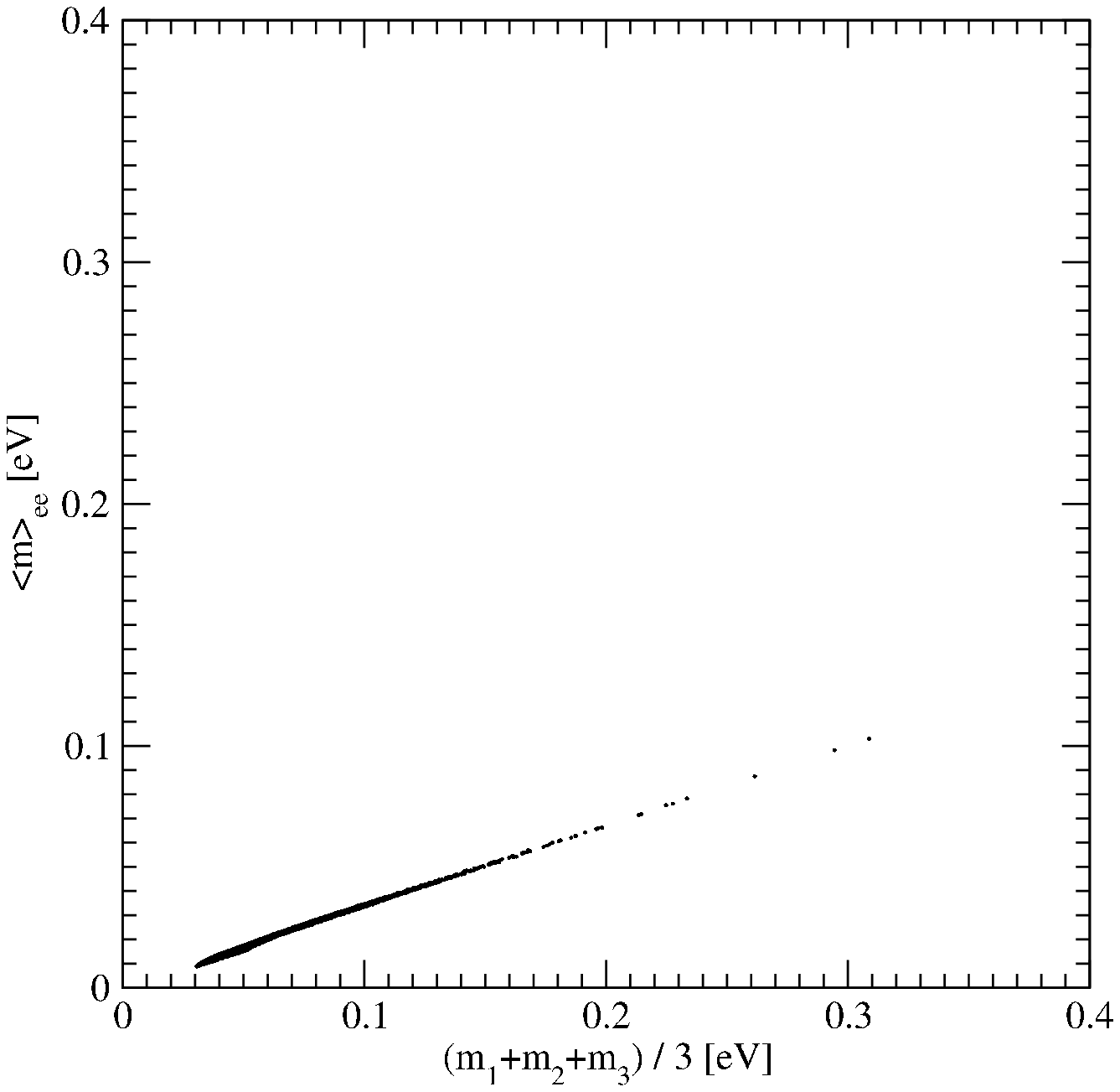}
\caption{The same as figure~7,
but in the case of an inverted neutrino mass spectrum.}
\label{fig8}
\end{figure}

\end{document}